\documentclass[10pt,notitlepage,nofootinbib,twocolumn]{revtex4-1}

\usepackage{graphicx,color,amsmath,amssymb}
\usepackage{bm} 
\usepackage{hyperref}

\begin{document}

\title{Universality of multiplicity distribution in proton-proton\\and electron-positron collisions}
\author{Adam Bzdak}
\email[E-Mail:]{bzdak@fis.agh.edu.pl}

\affiliation{AGH University of Science and Technology,\\
Faculty of Physics and Applied Computer Science,\\
30-059 Krak\'ow, Poland}

\begin{abstract}
It is argued that the multiplicity distribution in proton-proton ($pp$) collisions, which is often parameterized by the negative binomial distribution, may result from the multiplicity distribution measured in electron-positron ($e^{+}e^{-}$) collisions, once the fluctuating energy carried by two leading protons in $pp$ is taken into account.
\end{abstract}

\maketitle

\section{Introduction}
\label{sec: introduction}

The charged particle multiplicity distribution is one of the most basic
observables in high energy collisions. Although there is an abundance of
experimental results, see e.g., \cite{Kittel:2005fu,GrosseOetringhaus:2009kz}, on the
theory side this problem is poorly understood.

The multiplicity distribution measured in proton-proton ($pp$) collisions is
often parameterized by the negative binomial (NB) distribution \cite%
{Kittel:2005fu,GrosseOetringhaus:2009kz,Ansorge:1988kn,Breakstone:1983ns,Alexopoulos:1998bi}%
, which is characterized by two parameters: the mean number of particles $%
\left\langle n\right\rangle $, and $k$, which measures the deviation from
the Poisson distribution.\footnote{%
For NB $\langle n^{2}\rangle -\left\langle n\right\rangle ^{2}=\left\langle
n\right\rangle \left[ 1 + \frac{\left\langle n\right\rangle}{k} \right] $,
which goes to Poisson if $k\rightarrow \infty $ (at fixed $\left\langle
n\right\rangle $).} NB distribution works reasonably well, with certain limitations %
\cite{Kittel:2005fu,Szwed:1987vj}, 
for a broad range of energies and in total and limited phase-space rapidity
bins. For completeness we add that $k$ is a decreasing function of energy.

Interestingly, similar experimental observations were made in
electron-positron ($e^{+}e^{-}$) collisions, see e.g., \cite%
{GrosseOetringhaus:2009kz,Buskulic:1995xz}. NB works relatively well for
total and limited phase-space bins in rapidity and $k$ decreases with
energy. 

There are many similarities between $pp$ and $e^{+}e^{-}$, as far as the soft particle production is concerned,
but there are also
important differences. At the same $\sqrt{s}$, the mean number of particles
and $k$ are significantly larger in $e^{+}e^{-}$ than in $pp$.\footnote{%
For example, at $\sqrt{s}=200$ GeV in $pp$ collisions $k\approx 5$ (full
phase-space) in comparison to $k\approx 22$ in $e^{+}e^{-}$ at $\sqrt{s} \approx 100$
GeV, or $k\approx 16$ when extrapolated to $\sqrt{s}=200$ GeV.}

As pointed out in Refs. \cite{Basile:1980ap,Basile:1980by,Basile:1982we}
some differences between $pp$ and $e^{+}e^{-}$ can be easily understood. In $%
pp$ collisions a large fraction of initial energy, given by $\sqrt{s}$, is
carried away by two leading protons 
and is not available for particle production. This explains larger mean
multiplicity in $e^{+}e^{-}$, where the leading proton effect is not
present. This leads to the striking
relation between the total (full phase-space) mean number of charged particles in $pp$
and $e^{+}e^{-}$ interactions \cite{GrosseOetringhaus:2009kz}, see also \cite%
{Fermi:1950jd,Benecke:1975mg,Basile:1980by,Kadija:1993ie,Hoang:1994qz,Batista:1998ry,Chliapnikov:1990bc,
Sarkisyan:2010kb,Sarkisyan:2016dzo}: 
\begin{equation}
N_{\text{pp}}(\sqrt{s})=N_{\text{ee}}(K\sqrt{s})+2; \,\,\,\,\,\,\,\,\,\, K=0.35,  \label{mean_1}
\end{equation}%
that is, the mean number of particles in $pp$ at a given $\sqrt{s}$ is given
by the mean number of particles in $e^{+}e^{-}$ at $K\sqrt{s}$, plus two
leading protons. It turns out that the coefficient of inelasticity, $K$, present in Eq. (\ref{mean_1}) is approximately energy independent (see Fig. $10$ in Ref. \cite{GrosseOetringhaus:2009kz}) and Eq. (\ref{mean_1}) works surprisingly well from $30$ to $1800$ GeV. It remains
to be verified at the LHC energy. More recently, certain similarities
between $e^{+}e^{-}$ and ultra-relativistic heavy-ion collisions at RHIC and the LHC were
reported \cite{Back:2006yw,Sarkisyan:2016dzo}. See further discussion in section IV.

Equation (\ref{mean_1}) is suggestive of a universal mechanism of particle
production (or more precisely, a universal mechanism of hydronization) in both systems, controlled mainly by the actual energy deposited
into particle creation \cite{Basile:1980ap}. In $e^{+}e^{-}$ all initial
energy is consumed by produced particles, whereas in $pp$ the \textit{%
effective} energy available for particle production is given by
\begin{eqnarray}
E_{\text{eff}}^{2} &=&(p_{1}+p_{2}-q_{1}-q_{2})^{2}  \notag \\
&\approx &s\left( 1-x_{1}\right) (1-x_{2}),  \label{E_eff}
\end{eqnarray}%
where $p_{i}$ and $q_{i}$ are the incoming and the leading proton momenta,
respectively. $x_{i}$ is a fraction of the longitudinal momentum carried by
a leading proton, $x_{i}=q_{i,z}/p_{i,z}$, and $s=(p_{1}+p_{2})^{2}$.

We note that a universal hadronization mechanizm in $e^{+}e^{-}$ and $pp$ collisions is strongly supported by the success of the statistical hadronization model \cite{Becattini:1996gy,Becattini:2008tx,Becattini:2009fv}, which provides a very good description of hadronic multiplicities with a common hadronization temperature.

In this paper we show that Eq. (\ref{mean_1}) can be naturally extended to the whole
multiplicity distribution. In particular, we demonstrate that the broad multiplicity 
distributions measured in $pp$ collisions naturally result from relatively narrow multiplicity distributions observed in $e^{+}e^{-}$ interactions once the effective energy, $E_{\text{eff}}$, in $pp$ is properly taken into account.  

\section{Leading protons}
\label{sec:leading}

The problem of multiplicity distribution is naturally more complicated than
the mean number of particles; see, e.g., \cite{Basile:1984tc}. To proceed we need to specify the leading
proton $x$ distribution.\footnote{In other words, we assume that the energy deposited into particle production fluctuates from event to event.} We choose the beta distribution 
\begin{equation}
f(x)\propto x^{\lambda }\left( 1-x\right) ^{\mu }.  \label{f_x}
\end{equation}%
It is supported by rather limited experimental evidence \cite%
{Chapman:1973kn,Basile:1983kb}; however, it seems a natural first choice.
We note that our discussion is of qualitative character and certain refinements concerning Eq. (\ref{f_x}) are certainly possible.
Having (\ref{f_x}) we obtain%
\begin{equation}
\left\langle x\right\rangle =\frac{1+\lambda }{2+\lambda +\mu },  \label{x_m}
\end{equation}%
which is the average momentum fraction taken by a leading proton.

Next we would like to clarify how $f(x)$ is related to Eq. (\ref{mean_1}). 
We obtain\footnote{%
In Ref. \cite{Basile:1982we} it was found that the $x$'s of two leading protons
are uncorrelated.}%
\begin{equation}
N_{\text{pp}}(\sqrt{s})=\int f(x_{1})f(x_{2})N_{\text{ee}}\left( E_{\text{eff%
}}\right) dx_{1}dx_{2}+2.  \label{mean_2}
\end{equation}%
Taking%
\begin{equation}
N_{\text{ee}}(\sqrt{s})=a+b\cdot s^{\alpha },  \label{N_ee}
\end{equation}%
where $\alpha \approx 0.17$ \cite{GrosseOetringhaus:2009kz} (see Section 4 for further discussion) we arrive at 
\begin{equation}
N_{\text{pp}}(\sqrt{s})=N_{\text{ee}}\left( \left\langle \left( 1-x\right)
^{\alpha }\right\rangle ^{1/\alpha }\sqrt{s}\right) +2.
\end{equation}%
It means that $0.35$ from Eq. (\ref{mean_1}) is not related to $\left\langle
1-x\right\rangle $, as naively expected, but 
to $\left\langle \left( 1-x\right) ^{\alpha }\right\rangle ^{1/\alpha
} $. The latter can be calculated analytically leading to the following
equation%
\begin{equation}
\left\langle \left( 1-x\right) ^{\alpha }\right\rangle =\frac{\Gamma \left(
1+\alpha +\mu \right) \Gamma (2+\lambda +\mu )}{\Gamma \left( 1+\mu \right)
\Gamma \left( 2+\alpha +\lambda +\mu \right) }=0.35^{\alpha },
\label{eq_alpha}
\end{equation}%
which constrains possible parameters of the beta distribution. Assuming $%
\left\langle x\right\rangle =0.4$ \cite{Basile:1983kb}, see Eq. (\ref{x_m}),
we obtain $\lambda \simeq -0.8$ and $\mu =-0.7$, which fully determines $%
f(x) $.

\section{Calculations and results}
\label{sec:results}

\begin{figure*}[t]
\begin{center}
\includegraphics[scale=0.45]{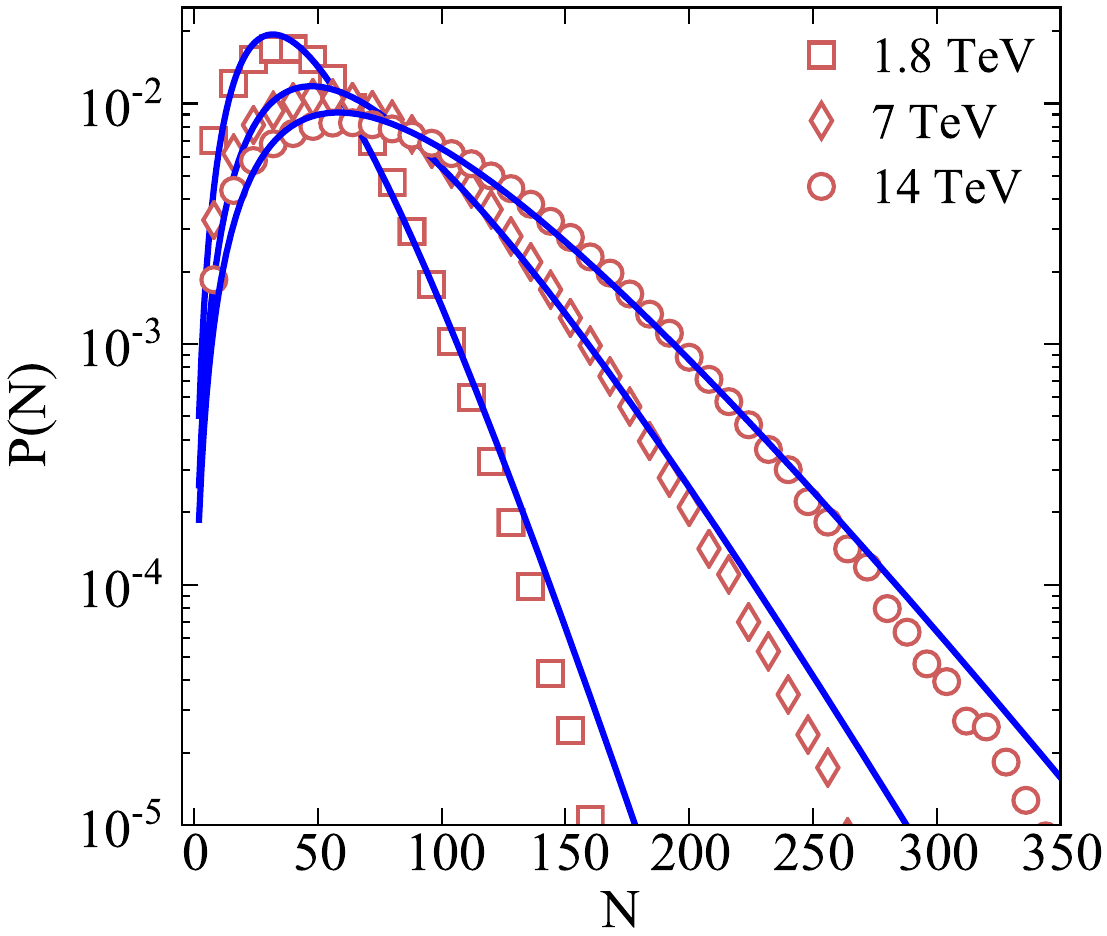} %
\includegraphics[scale=0.45]{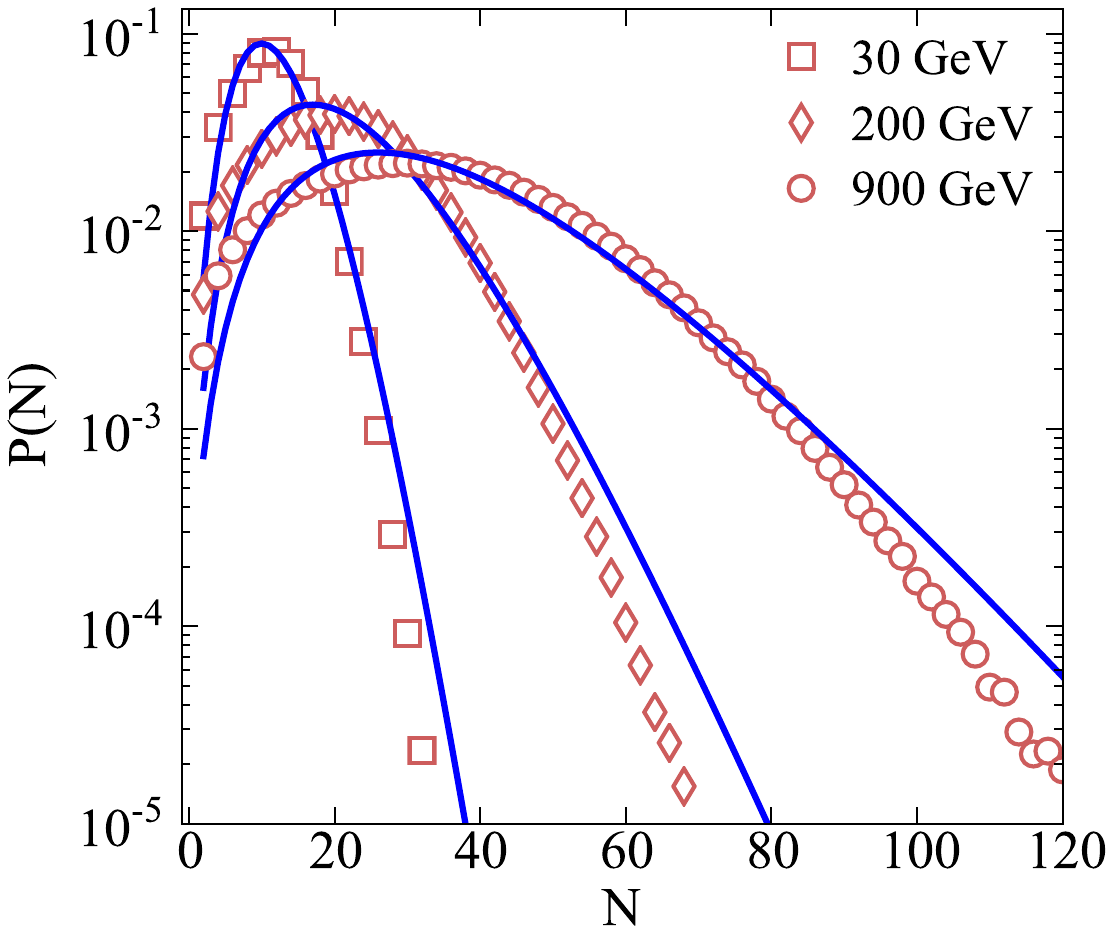}
\end{center}
\par
\vspace{-5mm}
\caption{Calculated full phase-space multiplicity distributions in proton-proton collisions
(open points) with NB distribution fits (lines). For clarity we show every
eighth (second) point in the left (right) plot.}
\label{fig:nbd}
\end{figure*}

The multiplicity distribution in $pp$ collisions, $P_{\text{pp}}(n)$, is
related to the multiplicity distribution in $e^{+}e^{-}$ interactions, $P_{%
\text{ee}}(n)$, as 
\begin{equation}
P_{\text{pp}}(n;\sqrt{s})=\int f(x_{1})f(x_{2})P_{\text{ee}}\left( n-2;E_{%
\text{eff}}\right) dx_{1}dx_{2},
\label{eq_Pn}
\end{equation}%
where $n\geq 2$. This equation is a straightforward generalization of Eq. (%
\ref{mean_2}). Instead of directly calculating the integral (\ref{eq_Pn})
we performed our calculations as follows.

First we sampled $x_{1}$ and $x_{2}$ of two leading protons from the beta
distribution, $f(x)$, with $\left\langle x\right\rangle =0.4$
and $\lambda =-0.8$, and calculated the effective energy\footnote{%
In our calculations we use the exact formula for $E_{\text{eff}}^{2}$ given
by $\left( \sqrt{s}-\sqrt{m^{2}+(x_{1}p_{z})^{2}}-\sqrt{%
m^{2}+(x_{2}p_{z})^{2}}\right) ^{2}-p_{z}^{2}\left( x_{1}-x_{2}\right) ^{2}$%
, where $p_{z}^{2}=s/4-m^{2}$ and $m$ is a proton mass.}, $E_{\text{eff}}$, available for particle
production in $pp$.\footnote{We accept only these events where $E_{\text{eff}}>0.3$ GeV so that at least
two pions can be produced.} Our choice of $\left\langle
x\right\rangle $ and $\lambda $ ensures that Eq. (\ref{mean_1}) is satisfied
with the right coefficient $0.35$. Next we sampled the number of particles from the multiplicity distribution
measured in $e^{+}e^{-}$ collisions at $\sqrt{s}=E_{\text{eff}}$. Clearly we do not know $P_{\text{ee}}(n)$ for all energies and thus we assume NB
with the mean given by Eq. (\ref{N_ee}), where $a=-2.65,$ $b=5.01$, $\alpha
=0.17$, and $k^{-1}=c+d\ln (\sqrt{s}),$ where $c=-0.066$ and $d=0.024$ \cite%
{GrosseOetringhaus:2009kz}.\footnote{%
Negative $k$ is rounded to the integer value. NB with a
negative integer $k$ becomes binomial distribution with the number of trials 
$-k$ and the Bernoulli success probability $-N_{\text{ee}}/k$. We also checked
that the Poisson distribution for $k<0$ leads to practically the same
results.} On top of that we add two particles corresponding to the leading
protons.

We performed our calculations at $\sqrt{s}=30,$ $200,$ $900,$ $1800,$ $7000$,
and $14000$ GeV. The results are shown in Fig. \ref{fig:nbd}, where the calculated 
full phase-space multiplicity distributions in $pp$ collisions (open symbols) are compared
with NB fits. We repeat that Eq. (\ref{mean_1}) is satisfied by
construction so our multiplicity distributions have the correct mean values,
see Tab. \ref{tab:2}. 
\begin{table}[b!]
\begin{center}
\begin{tabular}{|c|c|c|}
\hline
$\sqrt{s}$ [GeV] & $N_{\text{pp}}$ (model) & $N_{\text{pp}}$ (data) \\ \hline
$30.4$ & $11.4$ & $10.54\pm 0.14$ \\ \hline
$200$ & $21.1$ & $21.4\pm 0.6$ \\ \hline
$900$ & $34.6$ & $35.6\pm 1.1$ \\ \hline
$1800$ & $43.7$ & $45\pm 1.5$ \\ \hline
$7000$ & $69.1$ & -- \\ \hline
$14000$ & $87.5$ & -- \\ \hline
\end{tabular}%
\end{center}
\par
\vspace{-5mm}
\caption{Calculated mean number of charged particles in $pp$ collisions (full phase-space) compared with the experimental data.}
\label{tab:2}
\end{table}
The crucial test of our approach is the value of $k$, which we calculate as 
\begin{equation}
k=\frac{\left\langle N\right\rangle ^{2}}{\left\langle N^{2}\right\rangle
-\left\langle N\right\rangle ^{2}-\left\langle N\right\rangle },
\label{eq:k-nbd}
\end{equation}%
where $\left\langle N\right\rangle =N_{\text{pp}}$. In Tab. \ref{tab:1} we
list the obtained values of $k$ and compare them with available data. Taking into
account the simplicity of our approach, the agreement is satisfactory. 
\begin{table}[b!]
\begin{center}
\begin{tabular}{|c|c|c|}
\hline
$\sqrt{s}$ [GeV] & $k$ (model) & $k$ (data) \\ \hline
$30.4$ & $12.7$ & $9.2\pm 0.9$ \\ \hline
$200$ & $5.8$ & $4.8\pm 0.4$ \\ \hline
$900$ & $4.2$ & $3.7\pm 0.3$ \\ \hline
$1800$ & $3.8$ & $3.1\pm 0.1$ \\ \hline
$7000$ & $3.3$ & -- \\ \hline
$14000$ & $3.0$ & -- \\ \hline
\end{tabular}%
\end{center}
\par
\vspace{-5mm}
\caption{$k$ parameters, see Eq. (\ref{eq:k-nbd}), of calculated multiplicity distributions in $pp$
collisions compared with the experimental data.}
\label{tab:1}
\end{table}

\section{Discussion}

Several comments are in order.

(i) We do not offer any explanation of multiplicity distributions in $%
e^{+}e^{-}$ collisions. Our goal was to demonstrate that the problem of
multiplicity distributions in $pp$ could be reduced to $e^{+}e^{-}$ once the
fluctuating energy carried away by two leading protons in $pp$ collisions is taken into account.
We provided new evidence in favor of the hypothesis that the number of produced particles
in both systems (also possibly in heavy-ion collisions) is mostly driven by 
the amount of effective energy deposited into particle production, which 
naturally varies from event to event, and certain 
microscopic differences between the two systems are of lesser importance. 
In the literature this problem is extensively discussed in the context of the 
average number of particles. The fact that the similar connection holds between 
the widths of the full multiplicity distributions in $pp$ and $e^{+}e^{-}$ is new 
and not {\it a priori} expected.

(ii) In this paper we focused on the total phase-space multiplicity
distributions. It is plausible that the total
number of particles is determined (mostly) by the amount of available
energy. This is not obvious (expected) for limited phase-space bins since the
distribution of particles in transverse momentum or rapidity may be modified
by some nontrivial dynamics. This problem is much more difficult to tackle and 
any considerations would be strongly model dependent. For example, interesting 
collective effects were recently discovered in $pp$ collisions, see, 
e.g., \cite{Khachatryan:2010gv}, and their origin is still under debate \cite{Dusling:2015gta}. 
A possible parton rescattering (cascade, hydrodynamics) or other sources 
of correlations are not expected to significantly change the total number 
of produced particles.

(iii) The starting point of our analysis is the experimental 
observation summarized in Eq. (\ref{mean_1}). The coefficient of 
inelasticity, $K=0.35$, was found \cite{GrosseOetringhaus:2009kz} 
to be practically energy independent from $\sqrt{s}=30$ to $\sqrt{s}=1800$ GeV 
in contrast to certain dynamical models \cite{GrosseOetringhaus:2009kz}; 
see, e.g., Refs. \cite{Fowler:1988jz,Kadija:1993ie,Batista:1998ry}. 
The value of $K=0.35$ is often interpreted as a manifestation of the 
three-quark structure of the nucleon; see, e.g., Refs. \cite{Chliapnikov:1990bc,Hoang:1994qz,Sarkisyan:2010kb,Sarkisyan:2016dzo}. 
In a typical (minimum-bias) $pp$ collision 
roughly one constituent quark per nucleon interacts and this corresponds to 
an average inelasticity of $K\approx 1/3$. In heavy-ion collisions a nucleon 
usually undergoes more collisions and thus more quarks per nucleon are 
involved in particle production, leading to a higher value of $K$ \cite{Back:2006yw}.
In fact, the average number of particles produced in heavy-ion collisions is quite well 
described in a wounded quark or quark-diquark model; 
see, e.g., Refs. \cite{abab,Adler:2013aqf,Bozek:2016kpf,Lacey:2016hqy}. 
In this paper we argue that an event-by-event fluctuation of $K$ can naturally connect the 
multiplicity distributions measured in $e^{+}e^{-}$ and $pp$ collisions and it would be 
interesting to investigate the full multiplicity distributions in proton-nucleus ($pA$) and nucleus-nucleus ($AA$) 
collisions \cite{Akindinov:2007rr,Sarkisyan:2016dzo}.  

(iv) The main uncertainty of our approach is the leading proton $x$
distribution given in Eq. (\ref{f_x}). This form is partly supported by
existing data, but at rather limited energies and ranges of $x$. Thus it
should be treated as an educated guess, which hopefully is not far from reality. 
In addition, we assumed that 
$f(x)$ is energy independent, which is not proven experimentally. A mild
energy dependence is indicated by theoretical studies of \cite{Batista:1998ry}.
The agreement between the model and the data presented in Tab. \ref{tab:1} suggests 
that the assumed leading proton $x$ distribution might be an acceptable first approximation.

(v) To calculate the multiplicity distribution in $pp$ collisions one needs, as an input, 
the multiplicity distribution in $e^{+}e^{-}$ at all energies, see Eq. (\ref{eq_Pn}). 
In this paper we assumed that $e^{+}e^{-}$ follows a NB distribution (with the mean given 
by Eq. (\ref{N_ee}) and $k^{-1}$ discussed in Section III), which should be a reasonable 
approximation for our semi-quantitative study. As seen in Fig. \ref{fig:nbd}, the obtained 
multiplicity distributions in $pp$ collisions are close to NB with certain deviations. 
For example, the NB fits overestimate calculated
multiplicity distributions for higher values of $N$. Interestingly, similar
trends are seen in experimental data; see, e.g., Fig. $6$ in 
Ref. \cite{GrosseOetringhaus:2009kz} or Figs. $3-5$ in Ref. \cite{Ghosh:2012xh}.
Also it is known that for higher energies NB seems to fail for both $e^{+}e^{-}$ and $pp$ collisions \cite{GrosseOetringhaus:2009kz,Buskulic:1995xz}. This is not in contradiction to our study. 
In fact, if the multiplicity distribution is revealing a new structure at a given energy in $e^{+}e^{-}$, 
we expect the same phenomena to appear in $pp$ collisions but at different (higher) energies.
In this paper we focus on the width of the multiplicity distribution, being the first step after 
the mean number of particles, and thus detailed questions regarding an exact shape of 
the multiplicity distributions are not fully addressed in this paper.

(vi) We extrapolated $N_{\text{ee}}(\sqrt{s})$ into higher
energies using the 3NLO QCD result \cite{Dremin:2000ep,GrosseOetringhaus:2009kz}, which 
is almost identical to the NLO QCD fit (see, e.g., Fig. 10 in Ref. \cite{GrosseOetringhaus:2009kz}) 
and Eq. (\ref{N_ee}). For $k^{-1}$
we assumed it is a linear function of $\ln (\sqrt{s})$ up to the LHC energies.

(vii) There are many sophisticated Monte Carlo models 
(PYTHIA \cite{Sjostrand:2014zea}, HIJING \cite{Wang:1991hta}, EPOS \cite{Pierog:2009zt} etc.) 
that are used to describe the multiplicity distributions in various colliding systems.
However, we are not aware of any Monte Carlo model that would naturally explain Eq. (\ref{mean_1}), 
which, as discussed earlier, is usually interpreted in the constituent quark picture.
It would be very interesting to investigate this problem in detail, in particular, to see to what 
extent the multiplicity distribution in $pp$ is related to the multiplicity 
distribution in $e^{+}e^{-}$ interactions. 

(viii) 
The particle production in $pp$ and $AA$ collisions can be successfully described in the color glass condensate (CGC) approach; see, e.g., \cite{Gelis:2009wh,Tribedy:2010ab,Levin:2011hr}. In Ref. \cite{Gelis:2009wh} it was shown that the production of gluons from glasma color flux tubes follows the negative binomial distribution. In Ref. \cite{Tribedy:2010ab} the measured $pp$ multiplicity distributions were described within the CGC multi-particle production framework.
In Ref. \cite{Levin:2011hr}, the authors argue that the mean number of particles can be described with an input from  jet production in $e^{+}e^{-}$ annihilation. It would be interesting to see if the full multiplicity distribution in $pp$ can be described in a similar manner.

\bigskip

\section{Conclusions}
\label{sec:conclusions}

In conclusion, we argued that the full phase-space multiplicity distribution in $pp$
collisions is directly related to the multiplicity distribution in $%
e^{+}e^{-}$ interactions, once the leading proton effect in $pp$ is properly
accounted for. In $pp$ a large fraction of initial energy, roughly $1/2$ on
average, is carried away by two leading protons and is not available for
particle production. This component fluctuates from event to event, which
results in a significantly broader multiplicity distribution in $pp$ than in $e^{+}e^{-}$.
We provide a new argument in favor of a common mechanism of soft particle
production in both systems, which is mainly driven by the amount of
energy available for particle production.


\vspace{\baselineskip} 
\noindent\textbf{Acknowledgments} 
\newline
{} 
I thank Andrzej Bialas for useful discussions. 
Supported by the Ministry of Science and Higher Education (MNiSW) and 
by the National Science Centre, Grant No. DEC-2014/15/B/ST2/00175, and 
in part by DEC-2013/09/B/ST2/00497.

\end{document}